\documentclass{appolb}
\usepackage{graphicx}

\begin{document}
\title{The 3D structure of the Nucleon in momentum space: \\ 
TMD phenomenology%
\thanks{Presented at {\it Diffraction and Low-x}, 8-14 Sept. 2024, Hotel Tonnara Trabia, Palermo (Italy). Extended version of the proceedings to appear in Acta Physica Polonica B - Proc. Suppl.}%
}
\author{Marco Radici
\address{INFN - Sezione di Pavia, via Bassi 6, I27100 Pavia, Italy}
}
\maketitle
\begin{abstract}
I give a brief overview of our current understanding of the internal partonic 3D structure of nucleons in momentum space. I discuss some recent extractions of transverse-momentum-dependent distributions for quarks, whose analyses in the unpolarized case are reaching a theoretical precision comparable to collinear parton distribution functions. On the contrary, gluon transverse-momentum dependent distributions are poorly known from a phenomenological point of view. I briefly review their general properties and sketch a recent model calculation covering all (un)polarized combinations at leading twist. 
\end{abstract}
  
\section{Introduction}
\label{sec:intro}

\begin{figure}[htb]
\centerline{%
\includegraphics[width=13cm]{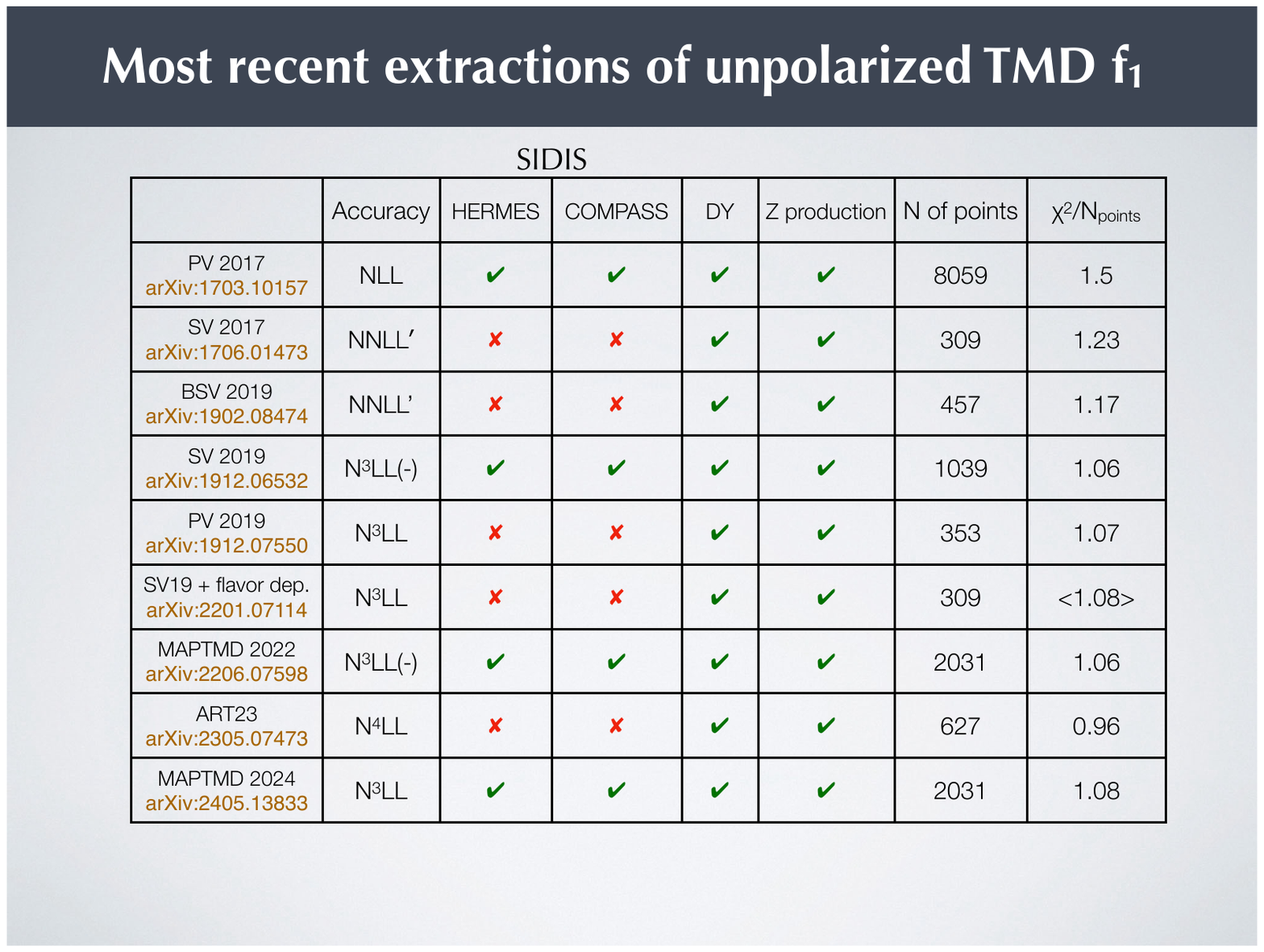}}
\caption{List of most recent extractions of unpolarized quark TMD PDF in unpolarized nucleon. From left to right, the second column "Accuracy" indicates the perturbative order in resummation of large logarithms, columns 3 and 4 refer to fitting SIDIS data from Hermes and Compass Collaborations, respectively, and columns 5 and 6 refer to Drell-Yan (DY) and Z-boson production data at fixed-target and collider facilities.}
\label{fig:list_f1}
\end{figure}

Multi-dimensional maps of the internal structure of hadrons are essential ingredients to understand the dynamics of confined colored objects within QCD. Such 3D maps in momentum space are called transverse-momentum-dependent parton distributions (TMD PDFs) and fragmentation functions (TMD FFs). At leading twist, eight different TMD PDFs exist for a quark in a nucleon depending on their polarization status (unpolarized, longitudinally or transversely polarized)~\cite{Mulders:1995dh,Boer:1997nt}, and similarly for TMD FFs describing the fragmentation of a quark into a hadron with spin 0 or $1/2$~\cite{Metz:2016swz}. Each of the eight TMD PDFs can be extracted from a measurable (spin) azimuthal asymmetry (for a review, see for example Refs.~\cite{Angeles-Martinez:2015sea,Albino:2008aa,Boussarie:2023izj}). The most important TMD PDF is the probability density of finding an unpolarized quark $q$ in an unpolarized hadron, $f_1^q$, because it enters the denominator of all asymmetries used to extract the other (polarized) TMD PDFs. The $f_1^q$ in the proton is also the best known TMD PDF; in Fig.~\ref{fig:list_f1}, its most recent extractions are listed (for the phenomenological extraction of $f_1^q$ in the pion, see Refs.~\cite{Wang:2017zym,Vladimirov:2019bfa,Cerutti:2022lmb,Barry:2023qqh}). 

In Fig.~\ref{fig:list_f1}, the column ``Accuracy" indicates the level of sophistication in resumming perturbative large logarithms of soft gluon radiation. From left to right, columns 3-6 indicate exclusion (red cross) or inclusion (green check mark) in the fit of Semi-Inclusive Deep-Inelastic Scattering (SIDIS) data from Hermes and Compass Collaborations, and of Drell-Yan (DY) and Z-boson production data from fixed-target and collider facilities, respectively (see the corresponding paper listed in the leftmost column for explicit reference to publications with experimental data used in the fit). The two rightmost columns show the quality of the fit to the included experimental data. By reading the table from top to bottom, it is easy to realize how recent extractions are produced from global analyses of large data sets (sometimes, very large like in Ref.~\cite{Bacchetta:2017gcc}), and at the same time reaching high perturbative accuracies comparable to extractions of collinear parton distribution functions (PDFs). 

A repository for TMD PDFs and FFs is available~\cite{Abdulov:2021ivr}, although currently it does not include some of the most recent extractions listed in Fig.~\ref{fig:list_f1}.

\section{Formalism}
\label{sec:formalism}

In the framework of TMD factorization~\cite{Collins:2011zzd}, the azimuthally symmetric differential cross section for the unpolarized DY process $h_A h_B \to \gamma^*/Z + X \to \ell^+ \ell^- + X$ can be written as~\cite{Bacchetta:2022awv}
\begin{equation}
\frac{d\sigma^{\mathrm{DY}}}{dq_T dy dQ} \propto x_A x_B H^{\mathrm{DY}} \sum_q c_q  \int_0^\infty \frac{d b_T}{2\pi} b_T J_0 (b_T q_T) \hat{f}_1^q (x_A, b_T^2; Q) \hat{f}_1^{\bar{q}} (x_B, b_T^2; Q) \, , 
\label{eq:DYcross}
\end{equation}
where $q_T, \, y$ and $Q$ are the transverse momentum, rapidity and invariant mass of the final lepton pair $(\ell^+ \ell^-)$, respectively, the $c_q (Q^2)$ is the electroweak charge of quark $q$, the hard factor $H^{\mathrm{DY}}(Q)$ admits a perturbative expansion and equals 1 at leading order (LO). The $\hat{f}_1^q$ is the Fourier transform of the TMD PDF of the unpolarized quark $q$ in hadron $h_A$, it depends on the quark longitudinal momentum fraction $x_A$ and on the variable $b_T$ Fourier-conjugated to the quark transverse momentum $k_T$ (similarly, for the antiquark $\bar{q}$ in hadron $h_B$ with momentum fraction $x_B$)~\footnote{In general, the TMD depends on the renormalization $\mu$ and rapidity $\zeta_A, \, \zeta_B$ scales; however, it's customary to choose $\mu^2 = \zeta_A = \zeta_B = Q^2$~\cite{Bacchetta:2022awv}.}. 

Similarly, for the SIDIS process $\ell N \to \ell + h + X$ on an unpolarized nucleon $N$ the azimuthally symmetric factorized cross section can be written as~\cite{Bacchetta:2022awv}
\begin{equation}
\frac{d\sigma^{\mathrm{SIDIS}}}{dx dz dq_T dQ} \propto x H^{\mathrm{SIDIS}} \sum_q e_q^2 \int_0^\infty \frac{d b_T}{2\pi} b_T J_0 (b_T q_T) \hat{f}_1^q (x, b_T^2; Q) \hat{D}_1^{q\to h} (z, b_T^2; Q) \, , 
\label{eq:SIDIScross}
\end{equation}
where $x, \, z$ are the usual SIDIS kinematic invariants, $Q$ is the invariant mass of the exchanged virtual photon with transverse momentum $q_T \sim - P_{hT}/z$, with $P_{hT}$ the transverse momentum of the detected final hadron $h$ produced by the fragmentation described by the TMD FF $D_1^{q\to h}$ (using the notation of transverse momenta from Ref.~\cite{Boer:2011fh}). 

The general structure of the TMD PDF in $b_T$-space reads
\begin{equation}
\hat{f}_1 (x, b_T^2; Q) = \mathrm{Evo}(Q \leftarrow \mu_{b_*}) \, [ C \otimes f_1] (x, \mu_{b_*}) \, f_{\mathrm{NP}} (x, b_T^2) \,  ,
\label{eq:f1}
\end{equation}
where the evolution operator Evo is perturbatively calculable. In order to avoid the appearance of large logarithms, the initial scale can be conveniently chosen as $\mu_{b_*} = 2 e^{-\gamma_E} / b_*$, with $\gamma_E$ the Euler constant~\cite{Collins:2011zzd}. The function $b_* (b_T)$ is such that at large $b_T$ it saturates to a fixed $b_{\mathrm{max}}$, thus avoiding that $\mu_{b_*}$ becomes too small and hits the Landau pole, thus making the TMD PDF perturbatively meaningful~\cite{Bacchetta:2017gcc,Bacchetta:2019sam,Bacchetta:2022awv,Bacchetta:2024qre}. This procedure introduces power corrections scaling like $(\Lambda_{\mathrm{QCD}}/q_T)^m$ with $m>0$~\cite{Catani:1996yz}, that for $q_T \approx \Lambda_{\mathrm{QCD}}$ need to be accounted for by the nonperturbative parametric term $f_{\mathrm{NP}}$, which must fulfill the constraint $f_{\mathrm{NP}} \to 1$ for $b_T \to 0$~\cite{Bacchetta:2019sam} and can be fitted to experimental data. 
In the perturbative small $b_T \ll 1/\Lambda_{\mathrm{QCD}}$ region, the TMD PDF 
can be matched onto the corresponding collinear PDF $f_1 (x)$ through the perturbatively calculable Wilson coefficients $C$. In this limit, the function $b_* (b_T)$ saturates to a fixed $b_{\mathrm{min}}$ value such that the final scale of the evolved TMD PDF is of the order of the hard scale $Q$~\cite{Bacchetta:2022awv}.

The ``Accuracy" column in Fig.~\ref{fig:list_f1} describes the perturbative accuracy reached in the description of $H^{\mathrm{DY}}$ in Eq.~(\ref{eq:DYcross}) and $H^{\mathrm{SIDIS}}$ in Eq.~(\ref{eq:SIDIScross}), of Evo and $C$ in Eq.~(\ref{eq:f1})~\cite{Bacchetta:2019sam}. The rightmost column in Fig.~\ref{fig:list_f1} describes the quality of the fit fixing the free parameters in $f_{\mathrm{NP}}$. Similar considerations hold for the TMD FF. 

For polarized TMD PDFs, an expression similar to Eq.~(\ref{eq:f1}) holds with the same universal Evo operator but with a different matching convolution and a different nonperturbative component, although the $b_*$ prescription for merging perturbative and nonperturbative regions should consistently be the same~\cite{Bacchetta:2020gko,Bacchetta:2024yzl}. For the case of the Sivers effect, this TMD PDF is the representative of the class of socalled na\"ive time-reversal odd (T-odd) functions~\cite{Boer:1997nt} that are non-universal but in a calculable way: in fact, based on very general assumptions like Lorentz invariance and color-gauge invariance QCD predicts that the Sivers TMD PDF extracted from the SIDIS process should have opposite sign to the one extracted from the DY process~\cite{Brodsky:2002cx,Ji:2002aa}. An experimental confirmation of this prediction is thus of paramount importance. Moreover, at small $x$ the quark Sivers TMD has been shown to be connected to the socalled spin Odderon~\cite{Dong:2018wsp,Kovchegov:2021iyc}, namely the C-odd 3-gluon exchange in the $t$-channel of the spin-dependent T-odd part of the dipole amplitude. The experimental confirmation of the inherited C-odd nature of the Sivers effect at small $x$ would be an indirect evidence of the existence of the spin Odderon. 

The polarized TMD PDFs and FFs dealing with transversely polarized quarks are called chiral-odd TMDs because they are connected to processes that flip the quark helicity~\cite{Jaffe:1996zw}. As such, they are suppressed in perturbative QCD~\cite{Kane:1978nd} but still can appear in cross sections at leading twist, provided that they are paired to another chiral odd partner. The most important representative is the transversity distribution $h_1^q$, which is the only leading-twist chiral-odd TMD PDF that exists also as a collinear PDF~\cite{Bacchetta:2006tn}. The transversity is very peculiar: 
in the nucleon, it evolves like a non-singlet function because it receives no contribution from gluons; its first Mellin moment, the socalled tensor charge $\delta q$, scales with the hard scale $Q$ (for a short review of $h_1$ properties, see Ref.~\cite{Radici:2024upx}). 
The transversity is also an important ingredient of effective field theories exploring new physics beyond the Standard Model, like new contributions of tensor operators to nucleon $\beta$-decay~\cite{Bhattacharya:2011qm} or to neutron permanent electric dipole moment~\cite{Dubbers:2011ns,Yamanaka:2017mef}.

\section{Selected results}

In the following, we give a brief overview of some recent and relevant results for the unpolarized TMD PDF $f_1$, the Sivers $f_{1T}^\perp$ and the transversity PDF $h_1$ and its related tensor charge.

\begin{figure}[htb]
\centerline{%
\includegraphics[width=13cm]{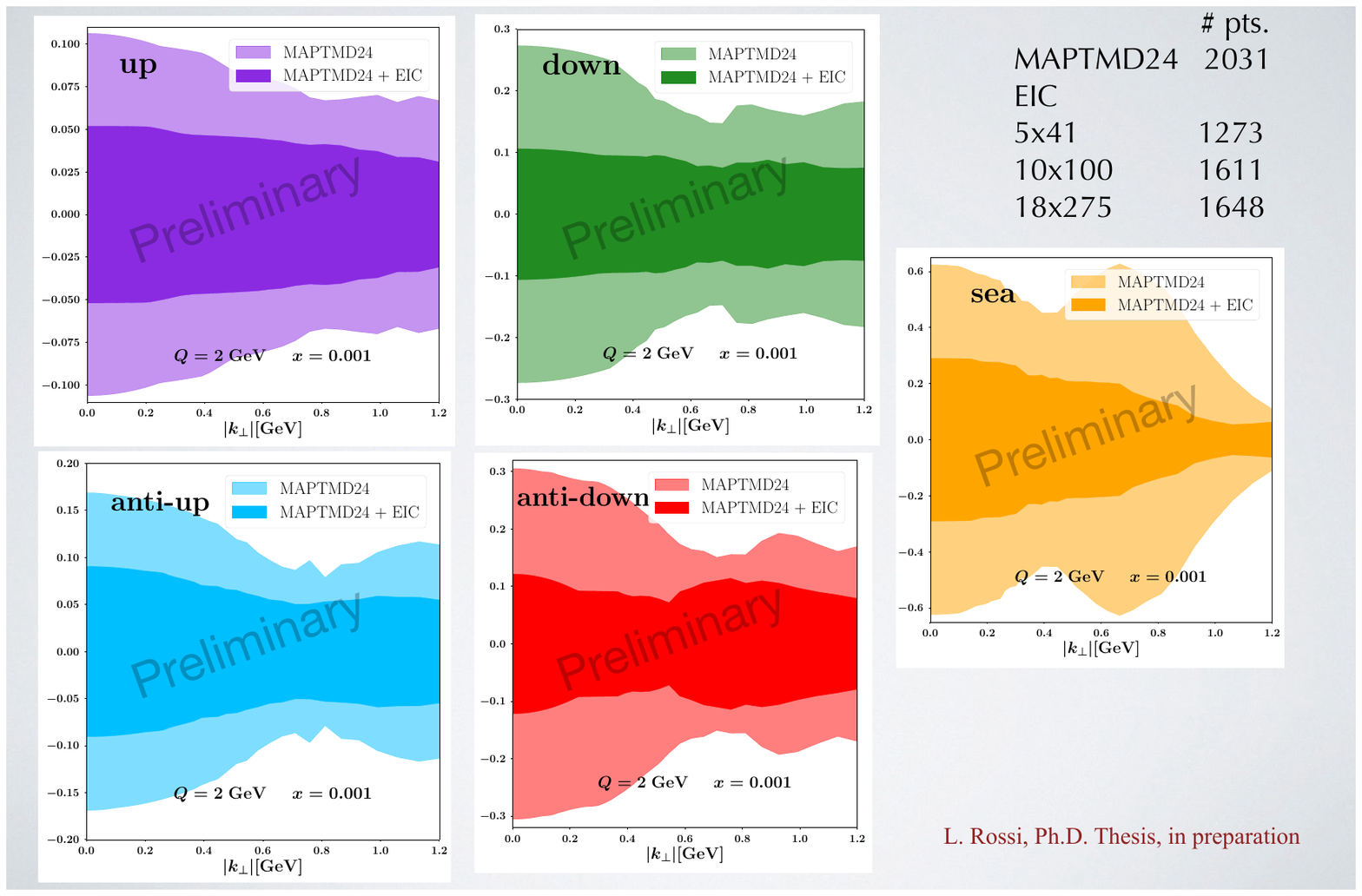}}
\caption{Impact on relative uncertainty $(f_1^q - \langle f_1^q \rangle) / \langle f_1^q \rangle$ for $q=u, d, \bar{u}, \bar{d}$, and a remaining cumulative sea quark contribution, as a function of $k_T$ at $x=0.001$ and $Q=2$ GeV, from EIC pseudodata at three different kinematics in the conditions of simulation campaign of May 2024. Baseline result from the MAPTMD24 extraction of Ref.~\cite{Bacchetta:2024qre}.}
\label{fig:EICimpact}
\end{figure}

\subsection{Unpolarized quark distribution}
\label{sec:f1}

The bottom line entry in the table of Fig.~\ref{fig:list_f1}, labelled MAPTMD24, points to the first ever global fit of SIDIS and DY data including flavor sensitivity to the quark intrinsic transverse momentum~\cite{Bacchetta:2024qre}. Using the extracted unpolarized TMD PDF $f_1^q$ as a baseline for $q = u, d, \bar{u}, \bar{d}$ and a cumulative sea quark contribution, in Fig.~\ref{fig:EICimpact} we show the impact by adding also pseudodata at three different kinematics of the future Electron-Ion Collider (EIC). The relative uncertainty $(f_1^q - \langle f_1^q \rangle) / \langle f_1^q \rangle$ at $x=0.001$ and $Q=2$ GeV is significantly reduced overall by a factor 2, with a remarkable reduction by a factor 3 for $d$ and $\bar{d}$. The capability of being sensitive to intrinsic $k_T$-distributions of different flavors is relevant also for the extraction of important Standard Model parameters like the $W$ boson mass, whose uncertainty should include also the contribution of this nonperturbative effect~\cite{Bacchetta:2018lna,Bozzi:2019vnl}.

\subsection{Sivers effect}
\label{sec:Sivers}

\begin{figure}[htb]
\centerline{%
\includegraphics[width=10.5cm]{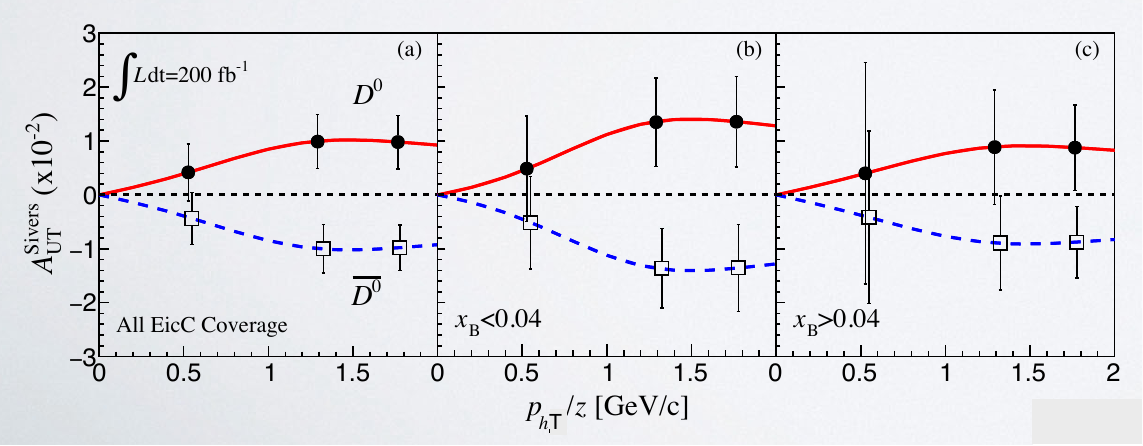}}
\caption{The Sivers asymmetry $A_{UT}$ for the $e + p^\uparrow \to D^0 / \bar{D}^0 + X$ process at the EIcC kinematics, as a function of the $D^0 / \bar{D}^0$ transverse momentum $P_{hT}/z$~\cite{Zhu:2024iwa}.}
\label{fig:SiversEIcC}
\end{figure}

The Sivers TMD PDF $f_{1T}^\perp$ represents the distortion of the $k_T$-distribution of an unpolarized quark induced by the transverse polarization of the parent nucleon~\cite{Sivers:1990cc}. Therefore, it is indirectly connected to the quark orbital angular momentum. It is also the prototype of the socalled na\"ive T-odd TMDs that vanish if there are no residual interactions between the active quark and the residual partons~\cite{Boer:1997nt}. As such, it is not a universal object and QCD predicts that $f_{1T}^\perp \vert_{\mathrm{SIDIS}} = - f_{1T}^\perp \vert_{\mathrm{DY}}$~\cite{Brodsky:2002cx,Ji:2002aa}. Several extractions of $f_{1T}^\perp$ are available in the literature (see, for example, the most recent Refs.~\cite{Cammarota:2020qcw,Bacchetta:2020gko,Echevarria:2020hpy,Bury:2021sue,Boglione:2021aha}~\footnote{We remind that the density of unpolarized quarks in a nucleon with transverse polarization ${\bf S}_T$, mass $M$ and momentum ${\bf P}$, reads $\rho^q = f_1^q - f_{1T}^{\perp\, q} \, ({\bf k}_T \times {\bf S}_T) \cdot \hat{\bf P}/M$ and demands for a consistent extraction of $f_1^q$ and $f_{1T}^{\perp\, q}$ with the same prescription for merging perturbative and nonperturbative regions; such kind of nucleon tomography was achieved for the first time in Ref.~\cite{Bacchetta:2020gko}.}) but the theoretical accuracy and the size of available data are much lower than for the unpolarized $f_1$. As a consequence, all extractions more or less agree on the description of the $x$-dependence of $f_{1T}^{\perp\,q}$ for valence $q = u, d$ (see Ref.~\cite{Bacchetta:2020gko} and references therein), but the sea quark components turn out very small and with large errors, and the $k_T$-distribution of all flavors is mostly unconstrained. In addition, earlier Monte Carlo simulations~\cite{Bianconi:2005yj,Bianconi:2006hc} and the scarce DY data for the Sivers effect~\cite{RHICSPIN:2023zxx,COMPASS:2023vqt} imply that only hints of the sign change prediction emerge while a statistical confirmation is not yet available. A large impact on reducing the Sivers uncertainty is predicted at the EIC~\cite{ATHENA:2022hxb,AbdulKhalek:2021gbh}. Moreover, a recent simulation of the Sivers asymmetry $A_{UT}$ for $D^0$ and $\bar{D}^0$ production in SIDIS processes at the kinematics of the planned Chinese EIC (EIcC)~\cite{Anderle:2021wcy} shows in Fig.~\ref{fig:SiversEIcC} that a full coverage of the available phase space could allow to statistically distinguish the two signals and test the C-odd nature of the charm Sivers effect, offering an indirect evidence of the existence of the spin Odderon~\cite{Zhu:2024iwa}.  

\subsection{Transversity distribution}
\label{sec:transversity}

\begin{figure}[htb]
\centerline{%
\includegraphics[width=5.3cm]{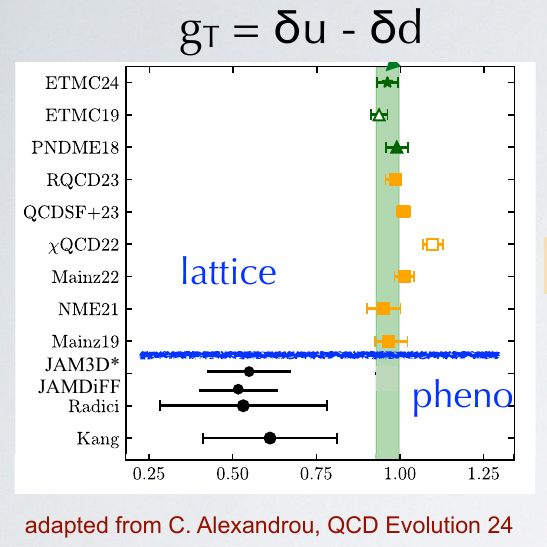} \hspace{0.2cm} \includegraphics[width=5cm]{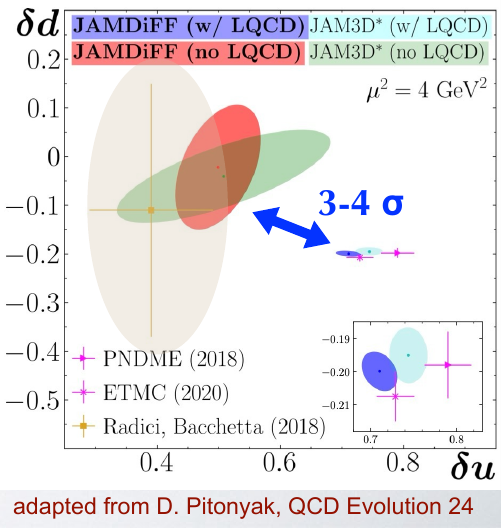}}
\caption{Left panel: isovector proton tensor charge $g_T = \delta u - \delta d$ at $Q=2$ GeV, computed in lattice and from phenomenological extractions of transversity ("JAM3D"~\cite{Gamberg:2022kdb} and "Kang"~\cite{Kang:2015msa} for Collins effect,  "JAMDiFF"~\cite{Cocuzza:2023vqs} and "Radici"~\cite{Radici:2018iag} for di-hadron mechanism). Right panel: $\delta d$ vs. $\delta u$ from lattice (magenta points~\cite{Gupta:2018qil,Alexandrou:2019brg}), di-hadron mechanism (yellowish ellipsis from Ref.~\cite{Radici:2018iag}, red and blue from Ref.~\cite{Cocuzza:2023vqs}, the latter including lattice points in the fit), Collins effect (green and light-blue ellipsis from Ref.~\cite{Gamberg:2022kdb}, the latter including lattice points in the fit).}
\label{fig:gT}
\end{figure}

The transversity distribution $h_1$ describes the net balance of quarks transversely polarized along or against the transverse polarization of the parent nucleon. Several channels are available to extract $h_1$ at leading twist as a TMD PDF or as a PDF, involving different unknown chiral-odd partner functions that must be independently determined from other processes, typically from semi-inclusive $e^+ e^-$ annihilations (for a short review, see Ref.~\cite{Radici:2024upx}). However, the transversity has been extracted so far only as a TMD PDF using the Collins effect~\cite{Collins:1993kk,Anselmino:2013vqa,Anselmino:2015sxa,Kang:2015msa,Gamberg:2022kdb}, or as a PDF using the inclusive production of di-hadron pairs~\cite{Collins:1994ax,Jaffe:1998hf,Radici:2001na,Bacchetta:2011ip,Bacchetta:2012ty,Radici:2015mwa,Radici:2018iag,Cocuzza:2023vqs}, where the collinear framework allows to include hadronic collision data from a rigorously proven factorization of the cross section~\cite{Radici:2016lam} (see also Ref.~\cite{Pisano:2015wnq} for a review)~\footnote{See also Ref.~\cite{Gliske:2014wba} for a unified description of single-hadron and di-hadron SIDIS cross section, and Ref.~\cite{Bacchetta:2023njc} for interesting connections between the di-hadron mechanism and the Collins effect for a leading hadron inside a jet~\cite{Kang:2017btw}.}. The related first Mellin moment, the tensor charge $\delta q$, can be precisely computed also on lattice, in particular the isovector component $g_T = \delta u - \delta d$ (for a review, see Ref.~\cite{Constantinou:2020hdm}), with resulting values that seem in tension with the extracted phenomenological results at $Q=2$ GeV, as shown in the left panel of Fig.~\ref{fig:gT}: results from Collins effect have labels "JAM3D"~\cite{Gamberg:2022kdb} and "Kang"~\cite{Kang:2015msa}; from di-hadron mechanism with labels "JAMDiFF"~\cite{Cocuzza:2023vqs} and "Radici"~\cite{Radici:2018iag}. In the right panel of Fig.~\ref{fig:gT}, the same message is delivered through the plot of $\delta d$ vs. $\delta u$. The yellowish~\cite{Radici:2018iag} and red~\cite{Cocuzza:2023vqs} ellipsis (denoting the results with the di-hadron mechanism) and the green ellipsis~\cite{Gamberg:2022kdb} (Collins effect) deviate by 3-4 standard deviations from the lattice results represented by the magenta points~\cite{Gupta:2018qil,Alexandrou:2019brg}. However, the authors of Refs.~\cite{Gamberg:2022kdb,Cocuzza:2023vqs} claim that compatibility can be reached by constraining the phenomenological fits to reproduce the lattice results, obtaining as final values of $\delta u$ and $\delta d$ the blue (di-hadron mechanism) and light-blue (Collins effect) ellipsis. Discussion is still open to scrutinize this result and resolve the apparent puzzle of few lattice points not altering the global $\chi^2$, thus being statistically irrelevant, whilst at the same time strongly influencing the fit and filling the original 3-4 $\sigma$ gap with phenomenological results. Future data from SoLID @JLab~\cite{Gamberg:2021lgx} and from the EIC~\cite{AbdulKhalek:2021gbh} will drastically improve the precision of phenomenological extractions, thus definitely clarifying if there is such phenomenology-lattice tension. 

\section{Gluon TMDs}
\label{sec:gluon}

\begin{figure}[htb]
\centerline{%
\includegraphics[width=4.5cm]{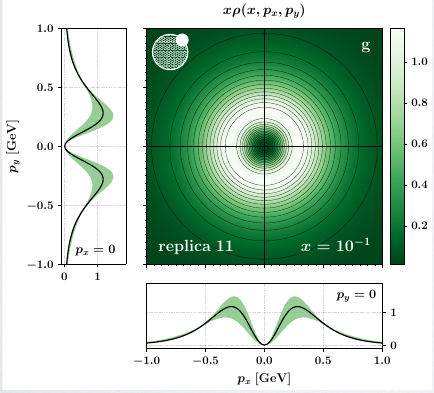} \hspace{0.2cm} \includegraphics[width=4.5cm]{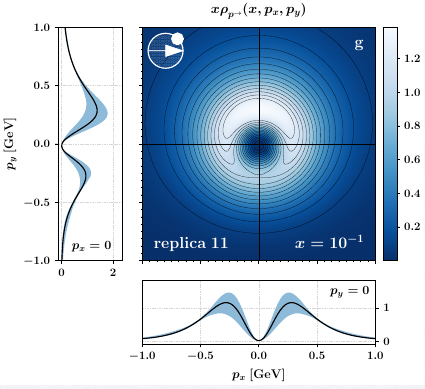} \hspace{0.2cm} \includegraphics[width=4.5cm]{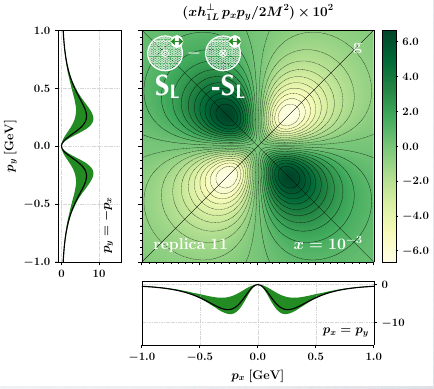}}
\caption{Left panel: density $x \rho^g = x f_1^g$ of unpolarized gluon in unpolarized proton, as function of $p_x, p_y$ at $x=0.1$~\cite{Bacchetta:2020vty}. Central panel: density $x \rho^g_{p^\to} = x f_1^g + p_y x f_{1T}^{\perp\, g} / M$ of unpolarized gluon in proton transversely polarized along $p_x$ at $x=0.1$.  Right panel: ``propeller" $h_{1L}^{\perp\, g}$ (scaled by $x p_x p_y / (2 M^2) \times 10^2$) for a linearly polarized gluon along $p_x$ at $x=0.001$ in a proton with longitudinal polarization $S_L$ perpendicular to the $(p_x, p_y)$ plane~\cite{Bacchetta:2024fci}. All plots at $Q_0 = 1.64$ GeV.}
\label{fig:gTMD}
\end{figure}

At leading twist, we have a table of eight different gluon TMD PDFs similar to the quark case, but now involving gluons unpolarized, circularly polarized or linearly polarized along a direction perpendicular to the momentum~\cite{Mulders:2000sh}. However, the class of T-odd gluon TMD PDFs is different from the quark one~\cite{Buffing:2013kca}. In fact, the more complicated color structure of the defining correlator makes the non-universality of gluon T-odd TMD PDFs more intricate: two distinct classes exist, the WW type and the dipole type, which describe different elementary mechanisms and, more importantly, happen in different processes (provided that factorization is proven). At small $x$, interesting links can be shown between gluon TMD PDFs and structures in the Color Glass Condensate~\cite{Dominguez:2010xd,Dominguez:2011wm}. In particular, T-odd gluon TMD PDFs of the WW type vanish while the ones of dipole type merge into the spin Odderon~\cite{Boer:2015pni}. While much is known of their theoretical properties, the experimental information on gluon TMDs is scarce and only few phenomenological studies are available (for a recent review focused on quarkonium production, see Ref.~\cite{Boer:2024ylx} and references therein). On the contrary, many model predictions have been released. As an example, we consider the spectator model of Refs.~\cite{Bacchetta:2020vty,Bacchetta:2024fci} where all leading-twist T-even and T-odd gluon TMD PDFs are calculated. In this model, the nucleon is represented as an active gluon plus a spectator on-shell spin-$1/2$ particle described by a parametric spectral function, whose parameters are fixed by reproducing NNPDF unpolarized and helicity gluon PDFs at $Q_0 = 1.64$ GeV~\cite{Bacchetta:2020vty} and T-odd structures are generated by computing gluon-spectator residual interactions at the one-gluon exchange level~\cite{Bacchetta:2024fci}. In Fig.~\ref{fig:gTMD}, from left to right, we show: 
\begin{itemize}
\item[-] the 2D plot in transverse momentum $(p_x, p_y)$ of the density of unpolarized gluon in unpolarized proton $x \rho^g = x f_1^g$ at $x=0.1$~\cite{Bacchetta:2020vty}, which is directly related to the T-even unpolarized $f_1^g$, 

\item[-] the 2D plot in transverse momentum $(p_x, p_y)$ of the density $x \rho^g_{p^\to} = x f_1^g + p_y x f_{1T}^{\perp\, g} / M$ of unpolarized gluon in proton transversely polarized along $p_x$ at $x=0.1$~\cite{Bacchetta:2024fci}, which involves $f_1^g$ and the T-odd Sivers function $f_{1T}^{\perp\, g}$, 

 \item[-] the ``propeller" function $h_{1L}^{\perp\, g}$ (scaled by $x p_x p_y / (2 M^2) \times 10^2$) for a linearly polarized gluon along $p_x$ at $x=0.001$ in a proton with longitudinal polarization $S_L$ perpendicular to the $(p_x, p_y)$ plane~\cite{Bacchetta:2024fci}. 
 \end{itemize}
 All plots calculated at the initial scale $Q_0 = 1.64$ GeV. 

\section{Summary}
\label{sec:end}

We can summarize the situation about TMD phenomenological studies as follows:
\begin{itemize}
\item[-] for unpolarized quark TMDs, we can assert that we entered a precision era: the TMD PDF $f_1^q$ for different flavors $q$ can now be extracted from high-quality fits of large data sets with high perturbative accuracy, comparable to state-of-art extractions of PDFs. Studies of the impact of this TMD PDF on observables at the LHC are now feasible, particularly for flavor-sensitive quantities like the $W$ boson mass parameter which is crucial for precision studies of Standard Model (SM) and beyond

\item[-] for polarized quark TMDs, our knowledge is more limited: the available data sets are smaller and the achieved perturbative accuracy in the theoretical formalism is lower. However, three major tasks can be pursued:
\begin{itemize}
\item[+] confirming the prediction of sign change between T-odd TMD PDFs extracted in SIDIS and DY processes, which is based on very general properties of QCD

\item[+] indirectly test the existence of the spin Odderon by looking at the properties of the Sivers TMD PDF at small $x$

\item[+] explore possible new physics beyond SM looking at chiral-odd structures like the transversity and its Mellin moment, the tensor charge
\end{itemize}

\item[-] for gluon TMDs, we have a complete classification at leading twist, including a detailed description of their process dependence and their evolution properties; factorization theorems have been proven for several processes of interest. However, while many models are available in the literature, only few phenomenological studies exist due to scarce experimental information. 
\end{itemize}

Important progresses and interesting developments are expected when more useful experimental data will become available, in particular from colliders like the planned EIC. 

\bibliographystyle{JHEP}
\bibliography{radici}

\end{document}